\documentstyle[aaspp4,psfig]{article}

\begin{document}

\title{Molecular Emission Lines from High Redshift AGNs\footnote{Invited
review to appear in ``Highly Redshifted Radio Lines,'' eds C. Carilli et al.}}
\author{Richard Barvainis}
\affil{MIT Haystack Observatory}

\begin{abstract}
A brief overview is given of the current status of molecular
emission line observations at high redshift.  New observations
are presented of CO in the gravitationally lensed quasar
MG 0414+0534 (detected) and the unlensed quasar 
PG 1634+706 (not detected).   Also noted are the results of 
a partially completed search for water maser emission in 
several redshift bands above $z=0.4$.
\end{abstract}


\section{Introduction}

Molecular lines at high redshift have generated considerable interest as
probes of the gaseous content of early galaxies, as signifiers of star
formation, and as indicators of AGN activation processes.  Thermal lines of
species like CO trace large masses of molecular hydrogen, typically $\sim
10^9 - 10^{11} M_{\sun}$, if detected in the distant universe.  According
to AGN activation scenarios developed over the past decade, it seems
likely that many quasars and active galaxies turn on when fuel in the
form of molecular gas is forced into a galactic nucleus during the
course of an interaction or merger between two galaxies.  This model
connects IR-luminous galaxies and classical quasars in an evolutionary
sequence that leads from concentration of gas in the nucleus through its
consumption or expulsion, eventually revealing a mature, optically bright
AGN (e.g.,  Sanders et al 1988; Barnes and Hernquist 1996).  During 
this sequence the compressed gas may pass through a luminous nuclear
starbursting phase.

Using molecular lines we hope to study these processes as they occured
in the early universe.  The observations push current instruments to
their limits, and consequently only a few sources have been detected at
high-$z$.  Several of these detections have been aided by a boost from 
gravitational lensing.  Below I summarize the observational situation
with regard to thermal (i.e., non-maser) molecular lines, including
discussion of a new detection of CO from a lensed quasar at $z = 2.64$.

I also briefly discuss an ongoing new search for H$_2$O masers at high
redshift.

\section{Thermal Lines}
\subsection{Low-$z$}
In the late 80's, with improvements in millimeter telescope sensitivities
and impetus from far-IR detections by IRAS, IR-luminous galaxies
and nearby AGNs began to be observed in molecular emission lines
[primarily CO(1--0)].  The AGNs detected included Mrk 231 (Sanders et al
1987), Mrk 1014 (Sanders, Scoville, \& Soifer 1988), I Zw 1 (Barvainis,
Alloin, \& Antonucci 1989), and a few other low-$z$ quasars (Sanders et
al 1989; Alloin et al 1992).  Revealing maps of the nucleus of NGC1068
in both CO and HCN have been made with millimeter-wave interferometers
(Tacconi et al 1994; Sternberg, Genzel, \& Tacconi 1994; Helfer \& Blitz
1995).  At moderate redshift, Scoville el al (1993) detected CO in the
radio galaxy 3C48 ($z=0.369$).  This measurement was interesting because
3C48 is a radio loud object which is clearly undergoing a merger that
has dumped copious molecular gas into an elliptical host (see Stockton
\& Ridgway 1992, and commentary by Barvainis 1993).

\subsection{High-$z$: The `old' standbys}
At high redshifts, the granddaddy of them all is of course IRAS
F10214+4724, at $z=2.28$, which has been detected in CO(3--2), CO(4--3),
and CO(6--5) (Brown \& Vanden Bout 1992; Solomon, Downes, \& Radford
1992), but probably not in CO(1--0) (detection reported by Tsuboi \& Nakai
1994 but not confirmed by Barvainis 1995).  A gravitationally lensed 
IR-galaxy/hidden-quasar, F10214 has now been imaged at high resolution in
CO, showing the $1.5''$ arc morphology seen in near-IR images (Downes,
this volume).  Various considerations suggest that the CO magnification
factor is $\sim 10$ (Downes, Solomon, \& Radford 1995).  The CO source
has been reported to be extended as well on scales of $4-5''$, well
beyond the boundaries of the infrared morphology (Scoville et al 1995).

The second gold-standard CO source at high-$z$ is H1413+117, the
Cloverleaf quasar (Barvainis et al 1994; Wilner, Zhang, \& Ho 1995),
another lensed object.  A total of 6 millimeter transitions at $z=2.558$
have been reported:  CO(3--2), CO(4--3), CO(5--4), CO(7--6),
CI($^3P_1-{^3P_0}$), and HCN(4--3) (Barvainis et al 1997).  Detailed
non-LTE modelling of the CO line strengths suggests that the molecular
gas is warm ($\ga 100$ K), dense ($n_{\rm H_2} \ga 3\times 10^3$
cm$^{-3}$), and not very optically thick ($\tau_{\rm CO} \la 3$).
Recent high-resolution ($0.5''$) imaging with the IRAM Plateau de Bure
Interferometer (PdBI) has clearly resolved the four optical spots in
CO(7--6) emission.  In the process of analyzing the CO images, evidence
for a cluster of galaxies at $z\sim 1.7$ was uncovered (Kneib et al
1997).  This cluster appears to be providing an extra boost in addition
to that of the (unseen) primary lensing galaxy.  The CO morphology of
the four spots is slightly different in the red and blue halves of the
line profile, and projecting back to the source plane suggests that the
CO is embedded in a rotating disk-like structure of diameter $\sim 200$
pc (Kneib et al 1997; Alloin, this volume).  The effective angular
resolution on the CO source using this procedure is $\sim 0.03''$, or
about 17 times smaller than the synthesized beam!  Yun et al (1997) also
modelled the morphology of the red and blue halves of the CO(7--6) line,
using $0.9''$ resolution OVRO images.  They arrived at a larger
structure ($\sim 1$ kpc) than that derived by Kneib et al (1997), having
a different position angle.  Why these two results differ is not known.

The third confirmed CO source is a quasar at $z=4.7$, BR1202--0125, which
is one of the highest redshift objects known.  Detected in CO(5--4) by
Ohta et al (1996) and Omont et al (1996), it was also detected in
CO(6--5) by Omont et al (1996).  The CO maps show two spots on the sky,
one at the position of the optical quasar, and another of equal strength
about $4''$ to the northeast.  There is no optical emission associated
with the northeast position in deep images (Hu, McMahon, \& Egami 1995).
Whether the northeastern CO spot is a lensed image of the quasar emission
or a dark companion galaxy is not known.  In the lensing hypothesis, the
northeast optical couterpart could be obscured by a patch of dust in the
lensing galaxy.

It is worth noting that all of the CO-detected objects mentioned above
also have far-IR (IRAS) or 1~mm/submm continuum detections, indicative of
highly luminous dust sources accompanying the CO-emitting gas.
Indeed, the presence of dust emission was the prime 
selection criterion for targeting CO searches in these objects, as it
has been for most other studies of distant CO. 

\subsection{High-$z$: Three newcomers} 
Three new high-$z$ CO systems have been reported recently.  All three
detections were made with interferometers (which are more reliable than
single dishes for measurement of weak, broad lines), but all, as of
this writing, have yet to be confirmed using another instrument or in
another CO transition.

The first is the $z=2.39$ radio galaxy 53W002, with a detection
of CO(3--2) using the OVRO interferometer reported by Scoville et al
(1997).  This object is interesting because it appears to lie within a
cluster of roughly 20 Ly-$\alpha$ emission line objects, the most
distant such cluster known (Pascarelle et al 1996).  Previously, a
possible detection of CO(1--0) had been reported from 53W002 by Yamada
et al (1995).  The Scoville et al result is particularly intriguing
because of a claimed extension of the CO source by about $3''$ or 15
kpc, with a velocity gradient along the major axis.  Such a large,
luminous CO source, if real, would be unprecedented -- all previously
known ultraluminous CO sources have been confined to the inner
kiloparsec or so of the host galaxy nucleus.  In contrast to the three
confirmed CO sources discussed above, 53W002 does not show any evidence for
gravitational lensing effects, nor has it been detected as a dust source
in the far-IR or submm continuum.

The second new source is BRI1335--0415, a quasar at $z=4.41$ detected in
CO(5--4) by Guilloteau et al (1997) using the PdBI.  This object
was previously detected as a dust source in the continuum at 1~mm; it
shows no {\it a priori} evidence for lensing.  The authors state that,
given the uncertainties in the CO-to-M(H$_2$) conversion factor, the
molecular mass in this system could be as high as $10^{11}$ M$_{\sun}$.
This can be compared to the case of the Cloverleaf where because of
lensing, and line modelling which suggests a low CO-to-M(H$_2$)
conversion factor, the estimated H$_2$ mass may be as low as a few
$\times 10^9$ M$_{\sun}$ (Barvainis et al 1997).  A more likely value
for BRI1335--0415 is M(H$_2$) $\sim$ a few $\times 10^{10}$, using the
conversion factor appropriate for ultraluminous IR galaxies (Solomon et
al 1997) and assuming no lensing boost.

\begin{figure}
\psfig{figure={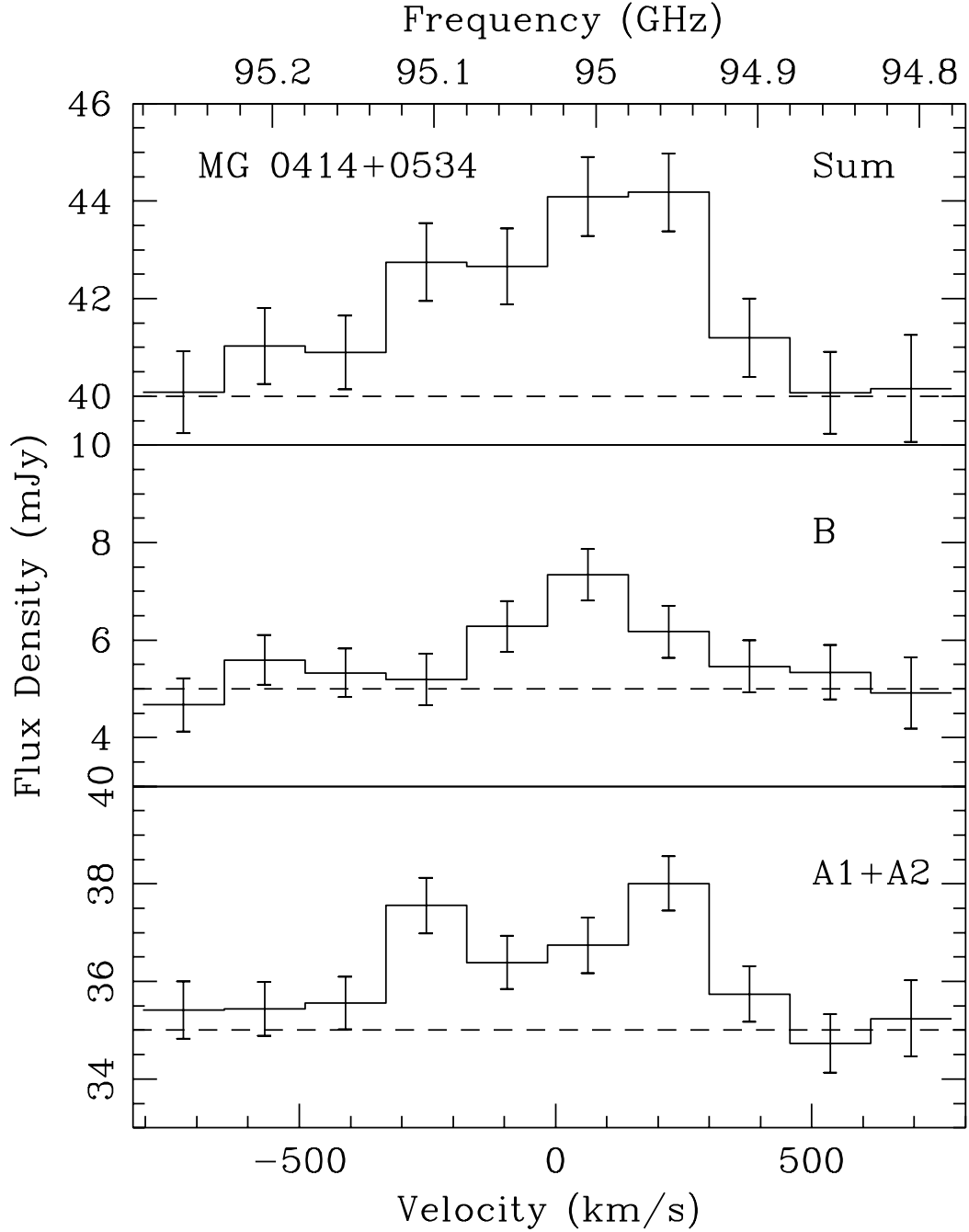}}
\caption{ CO($3-2$) spectra toward MG 0414+0534, smoothed
to a resolution of 158 km/s.  The velocity offsets are relative to a
redshift of 2.639.  The lower panel shows the spectrum of lensed
component A (= A1+A2), the middle panel component B, and the top panel
the sum.  } 
\end{figure}

The third new detection of CO at high redshift is from MG 0414+0534, a
lensed quasar at $z= 2.64$.  In the centimeter radio continuum it has
four components, with a maximum separation of $2''$ (Hewitt et al 1992).
We detected the CO(3--2) line at 95 GHz with high SNR using the PdBI
(Barvainis et al 1998).  The line is quite broad, at $\Delta v_{\rm FWHM}
= 580$ km s$^{-1}$, which is about as broad as the very broadest lines
seen in low redshift IR-luminous galaxies.  Although
the synthesized beam size of $2''$ was comparable to the separation
between the A (= A1+A2) and B radio images, it proved possible to
produce individual CO spectra for these components by fitting in the UV
plane.  These spectra, plus their sum, are shown in Figure 1.  When
higher SNR and higher resolution observations become available,
it may be possible to play the same trick with MG0414+0534 as Kneib et
al (1997) did with the Cloverleaf:  determining the very small-scale
kinematic structure of the CO source by deprojection, using a reasonably
accurate model for the lensing potential (such as that of Falco,
Leh\'ar, \& Shapiro 1997).

Finally, I report here a non-detection (see also Barvainis et al 1998).
The radio quiet object PG 1634+706 is an optically luminous quasar at
$z=1.33$, with detections in all four IRAS wavebands.  It is the most
distant unlensed IRAS source, and among the most luminous.  Its
spectral energy distribution in the IR/submm is very similar to those of
the Cloverleaf and F10214+4724, so PG 1634+706 seemed like a very good
candidate for CO detections.  We used the adjusted systemic redshift
estimate of Tytler \& Fan (1992), $z=1.337$, for the CO(2--1)
observations.\footnote{A recent measurement of PG 1634+706 in
[OIII]~$\lambda$5007 by Nishihara et al (1997) finds a systemic redshift
of $z = 1.336$;  therefore the redshift used for the CO(2--1) observations
differed from true systemic by only $\Delta z = 0.001$, or 128 km s$^{-1}$.}
No emission was detected, at a level somewhat below that expected (by a
factor of 2--4) if the CO-to-IR ratio were similar to that of other
distant CO sources.  

\section{Water masers}

In a search for very distant maser sources, a program has been started
at the Effelsberg 100m telescope (collaborators:  C.  Henkel, R.
Antonucci, S.  Baum, A.  Koekemoer) to try to detect H$_2$O masers in
several high redshift bands (corresponding to the receiver frequency
ranges available at the telescope).  There is reason to believe that
water maser power may scale approximately with x-ray luminosity
(Neufeld, Maloney, \& Conger 1994), making high-$z$ detections feasible.
The interest in high-$z$ water masers centers on their use as probes of
the inner nuclear regions in luminous AGNs, and on their potential
for making direct measurements of the source distance (as has been done
for NGC 4258; see contribution by Herrnstein, this volume).  At moderate
redshifts, such distance measurements could be turned into direct
estimates of $H_0$ independent of the usual distance ladder.  

So far we have observed a total of 60 objects in the redshift ranges
$0.46 < z < 0.55$, $1.46 < z < 1.79$, and $3.35 < z < 3.83$.  No
detections have yet been obtained, with typical maser line upper limits
of 20 mJy.  Observations of 40--50 more sources in other redshift ranges
are planned for a future observing run.

\end{document}